\documentclass[twocolumn,aps,prl,floatfix,superscriptaddress]{revtex4-1}

\usepackage{amsmath,amssymb}
\usepackage{graphicx,float}
\usepackage{times,txfonts}
\usepackage{color}
\usepackage{multirow}
\usepackage{bbold}
\usepackage{color}
\usepackage{soul}
\usepackage{hyperref}
\usepackage{times,txfonts}
\usepackage{natbib}
\usepackage{nicefrac}
\usepackage{blindtext}
\usepackage[export]{adjustbox}
\usepackage{mathrsfs}
\usepackage{textcomp}
\usepackage{multirow}
\usepackage[normalem]{ulem}
\usepackage{amsmath}
\newcommand{\bra}[1]{\left\langle #1 \right|}
\newcommand{\ket}[1]{\left| #1 \right\rangle}

\renewcommand{\epsilon}{\varepsilon}

\def\VR{\kern-\arraycolsep\strut\vrule &\kern-\arraycolsep}
\def\vr{\kern-\arraycolsep & \kern-\arraycolsep}
\usepackage{soul}
\definecolor{lightblue}{RGB}{185,210,248}
\definecolor{lgreen}{RGB}{15,150,15}
\sethlcolor{lightblue}

\begin{document}

\title{Three-fold coincidence by stimulated parametric down-conversion}

\author{Yishai Klein}
\affiliation{Nexus for Quantum Technologies, University of Ottawa, Ottawa, ON, K1N 5N6, Canada}

\author{Alessio D'Errico}
\affiliation{Nexus for Quantum Technologies, University of Ottawa, Ottawa, ON, K1N 5N6, Canada}

\author{Farid Ghobadi}
\affiliation{Nexus for Quantum Technologies, University of Ottawa, Ottawa, ON, K1N 5N6, Canada}

\author{Dilip Paneru}
\affiliation{Nexus for Quantum Technologies, University of Ottawa, Ottawa, ON, K1N 5N6, Canada}
\affiliation{Department of Physics, University of Naples Federico II, Naples, 80126, Italy}

\author{Nazanin Dehghan}
\affiliation{Nexus for Quantum Technologies, University of Ottawa, Ottawa, ON, K1N 5N6, Canada}

\author{Eliahu Cohen}
\affiliation{Faculty of Engineering and Institute of Nanotechnology and Advanced Materials, Bar-Ilan University, Ramat Gan, 52900, Israel}
\affiliation{Institute for Quantum Studies, Chapman University, Orange, California 92866, USA}

\author{Ebrahim Karimi}
\affiliation{Nexus for Quantum Technologies, University of Ottawa, Ottawa, ON, K1N 5N6, Canada}
\affiliation{Institute for Quantum Studies, Chapman University, Orange, California 92866, USA}

\begin{abstract}
Parametric down-conversion is a widely used source of nonclassical light in quantum optics and photonic quantum technologies. While stimulated parametric down-conversion with strong classical seeds is well studied, the regime in which stimulation occurs at the single-photon level has hitherto remained largely unexplored experimentally. Here, we study continuous-wave, low-gain down-conversion seeded by a weak coherent field with an average photon number well below one per coherence time. By measuring third-order temporal correlations, we observe a clear enhancement that cannot be accounted for by spontaneous processes or accidental coincidences alone, and is consistent with stimulation involving the seed and the generated photon pair. These results provide time-domain evidence of seed-induced three-photon correlations and suggest new ways to engineer and probe multi-photon states for quantum imaging, sensing, and information processing.
\end{abstract}
\maketitle

%==========================================================
\section{Introduction}
%==========================================================
Parametric down-conversion (PDC) is a nonlinear optical process in which a pump photon is converted into a pair of lower-energy photons, known as the \textit{signal} and \textit{idler}, through the second-order susceptibility of a $\chi^{(2)}$ medium~\cite{BurnhamWeinberg1970,HongMandel1985,Klyshko1988,ShihAlley1988}. In the spontaneous parametric down-conversion (SPDC) regime, this process is driven by vacuum \emph{fluctuations} and has become one of the most widely used sources of nonclassical light in quantum optics. Over the past decades, SPDC has enabled a range of foundational experiments, including Hong--Ou--Mandel interference~\cite{HongOuMandel1987,bouchard2021two}, the generation of polarization-entangled photon pairs~\cite{Kwiat1995}, and many key protocols in quantum information science~\cite{GisinThew2007,Eisaman2011,Pan2012Review}. Today, it remains a central platform for photonic quantum technologies, supporting applications that span quantum imaging~\cite{defienne2024advances} as well as quantum communication and computing~\cite{couteau2023applications}.\newline
Beyond the spontaneous regime, stimulated parametric down-conversion (StPDC) has been extensively studied within the broader framework of optical parametric amplification (OPA)~\cite{Klyshko1988,Louisell1961,Yariv1975,Caves1982,Boyd2008}. In these implementations, a coherent seed field is injected into one of the down-conversion modes, stimulating the emission of its conjugate partner and thereby enhancing the overall pair-production rate. High-gain realizations of this process have enabled the exploration of a range of phenomena, including micro--macro entanglement~\cite{DeMartini1998,ghobadi2013creating}, quantum cloning~\cite{SciarrinoDeMartini2005}, and macroscopic quantum superposition states~\cite{FerraroOlivaresParis2005}. Related techniques, such as stimulated emission tomography (SET), rely on the same mechanism to reconstruct biphoton states from classical measurements~\cite{LiscidiniSipe2013}. In all of these approaches, however, the seed field is classical, i.e. coherent state, typically containing many photons (with Poissonian statistics) per coherence time. In this regime, stimulated emission scales linearly with the seed amplitude, and the observed intensities and correlations can often be described within a classical nonlinear-optics framework, even though the underlying fields may exhibit quantum features such as squeezing or entanglement.

A different route to generating nonclassical light using StPDC is based on single-photon--added coherent states (SPACS)~\cite{AgarwalTara1991,lvovsky2001quantum,Zavatta2004}. In these schemes, a coherent state is injected into a low-gain SPDC process, and the detection of a photon in the conjugate mode heralds the addition of a single photon to the input field. The resulting states are inherently nonclassical and have provided a useful platform for exploring the gradual transition between classical coherent states and genuinely quantum states of light~\cite{Zavatta2004,zavatta2005single,parigi2007probing}. 

Despite these advances, an experimentally distinct regime of stimulated parametric down-conversion has remained largely unexplored. Specifically, this is the regime in which the seed is an ultra-weak coherent field containing, on average, far less than one photon per coherence time at the input of the nonlinear interaction. In this limit, the $\chi^{(2)}$ process directly couples individual input photons to the signal--idler modes. Unlike classical-seed OPA or SPACS-based schemes, where nonclassicality arises from amplification or conditional measurements, here it arises intrinsically from the stimulation process itself, at the single-photon level.

\begin{figure}[h!]
    \centering
    \includegraphics[width=0.9\linewidth]{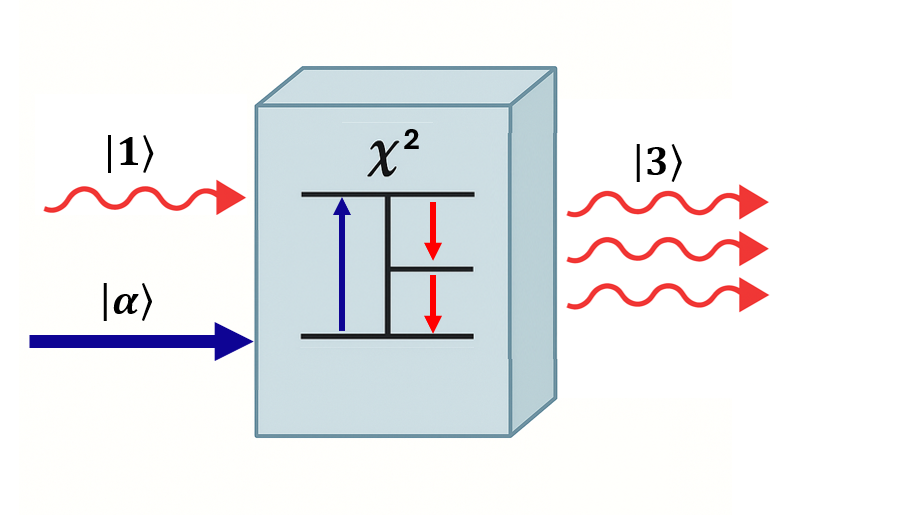}
    \caption{Concept of stimulated photon-triplet generation in a $\chi^{(2)}$ medium. A photon injected into one of the down-conversion modes stimulates the emission of a signal--idler pair from the pump. The seed photon and the generated pair are produced within the same temporal window, giving rise to a correlated three-photon event.}
    \label{fig:scheme}
\end{figure}

In this work, we investigate stimulated parametric down-conversion in the weak-seed regime, where a mode-matched coherent field with an average photon number well below one per coherence time is injected into the nonlinear interaction. In this limit, a single seed photon can stimulate the emission of a signal--idler pair, such that all three photons are generated within the same temporal window, forming a correlated triplet (Fig.~\ref{fig:scheme}). 
In a frequency-degenerate, effectively single-mode configuration, this process gives rise to a triplet contribution in the output field. Because the photons are indistinguishable after the interaction, we access this sector indirectly by splitting the field into three detection channels and measuring the third-order temporal correlation function. We observe a localized enhancement that appears only in the presence of the weak seed and cannot be explained by spontaneous emission alone, consistent with seed-induced three-photon correlations arising from stimulated down-conversion.

%%%%%%%%%%
%%%%%%%%%%

Higher-order correlations in parametric down-conversion have previously been explored by enhancing multi-pair emission, either through strong classical seeding or operation at higher gain, enabling measurable four-fold coincidence events~\cite{ou1999observation,lamas2001stimulated,eisenberg2005multiphoton}. Alternatively, cascaded $\chi^{(2)}$ processes have been used to generate photon triplets through sequential down-conversion stages~\cite{Hubel2010,Shalm2013}. In contrast, the approach considered here probes a different regime, in which an ultra-weak coherent seed participates directly in the nonlinear interaction at the level of individual photons. Rather than relying on higher-order emission or cascaded processes, this mechanism gives rise to a temporally localized three-photon signature in $g^{(3)}$. Our results show that the presence of a weak seed modifies the third-order temporal correlation structure of PDC in a way that is consistent with seed-induced three-photon correlations. This provides time-domain evidence of stimulated effects at the single-photon level and illustrates how the transition from spontaneous to stimulated emission can be accessed through higher-order correlation measurements. More broadly, these findings suggest new strategies for generating and probing multi-photon states in nonlinear optical systems, with potential applications in correlation-based imaging, sensing, and precision measurements that extend beyond second-order coherence.

%==========================================================
\section{Theoretical model for weak-seed PDC}
%==========================================================
In this section, we outline a simple model to describe two- and three-fold coincidence rates in parametric down-conversion in the presence of a weak seed. We consider a single spatio-temporal mode and focus on the frequency-degenerate case. Although simplified, this model captures the essential scaling of the three-fold signal with the seed amplitude.

\subsection{Unseeded SPDC: single-mode approximation}

We begin with the standard description of parametric down-conversion as a $\chi^{(2)}$ interaction in the undepleted pump approximation~\cite{gerry_knight}. In this approximation, the pump mode is taken to be in a strong coherent state $\ket{\alpha}$ and is treated as a classical field, so that its amplitude is absorbed into the effective nonlinear coupling. For type-I, frequency-degenerate SPDC, the interaction Hamiltonian in the interaction picture can then be written as
%%%%
\begin{equation}
	\hat H_I = \hbar \eta\,\hat a^\dagger \hat a^\dagger + \text{H.c.},
\end{equation}
%%%%%
where $\hbar$ is the reduced Planck constant, $\hat a^\dagger$ and $\hat a$ are the creation and annihilation operators of the degenerate mode, and $\eta \propto \chi^{(2)} \alpha$ characterizes the strength of the nonlinear interaction driven by the pump field.\newline
Starting from the vacuum state $\ket{0}$, the system evolves under the unitary operator $\hat U(t)=\exp(-i\hat H_I t/\hbar)$. In the low-gain regime, $|\eta t|\ll 1$, the output state can be approximated as
%%%%%
\begin{equation}\label{eq:SPDC_single_mode}
 	\ket{\Psi_{SPDC}}=\hat{U}(t)\,\ket{0} \simeq \ket{0} + \gamma \sqrt{2}\,\ket{2},
\end{equation}
%%%%%
where $\ket{n}=(\hat{a}^{\dagger})^n\ket{0}/\sqrt{n!}$ denotes an $n$-photon Fock state of the down-converted mode, and $\gamma$ is a small dimensionless parameter set by the pump amplitude and interaction time. In what follows, we retain only the non-vacuum component, as it is this contribution that gives rise to the measured coincidence signals.

The normally ordered second-order Glauber correlation function, with $\hat n=\hat a^\dagger \hat a$ the photon-number operator, is defined as
%%%%%
\begin{equation}\label{eq:Glauber}
	G^{(2)}(0)=\bra{\Psi_{SPDC}}:\!\hat{n}^2\!:\ket{\Psi_{SPDC}}=\bra{\Psi_{SPDC}}\hat{a}^\dagger \hat{a}^\dagger \hat{a} \hat{a}\ket{\Psi_{SPDC}},
\end{equation}
%%%%%
which in the low-gain limit scales as
%%%%%
\begin{equation}
    G^{(2)}(0) \propto 4|\gamma|^2 .
\end{equation}
%%%%%
%%%%%%%%%
\begin{figure*}[t]
    \centering
    \includegraphics[width=1\linewidth]{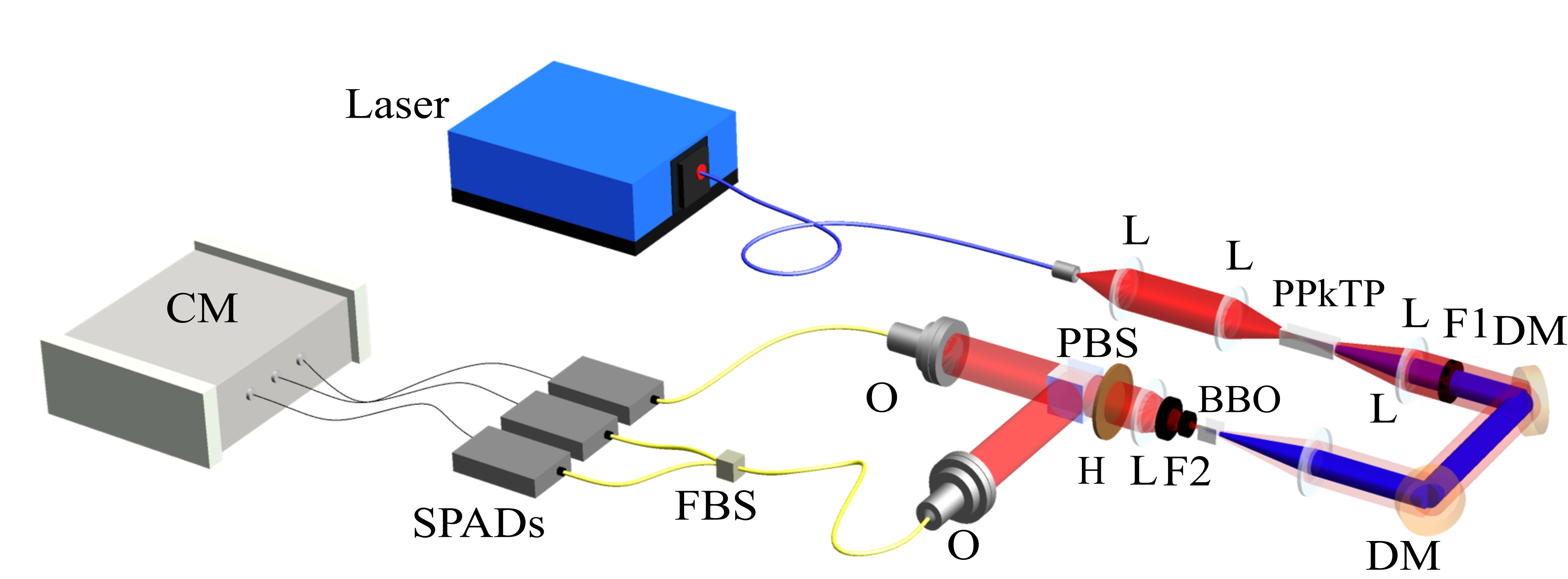}
    \caption{Experimental setup for weakly seeded stimulated parametric down-conversion. A continuous-wave laser at $800\,\mathrm{nm}$ is frequency doubled in a PPKTP cavity (custom-made by RAICOL) to generate a $400\,\mathrm{nm}$ pump for type-I SPDC in a BBO crystal. The residual fundamental beam is attenuated and mode-matched to provide a weak coherent seed. The down-converted photons and the seed are spectrally filtered, split into multiple channels, and coupled into single-mode fibers for detection with single-photon avalanche diodes (SPADs) and time-tagging electronics. F1: short-pass filter; F2: long-pass and band-pass filters; L: lens; DM: dichroic mirror; O: objective lens; PBS: polarizing beam splitter; FBS: fiber beam splitter; H: half-wave plate; CM: coincidence electronics.}
    \label{fig:setup}
\end{figure*}
%%%%%%%%%

\subsection{Adding a weak seed}
We now consider the same interaction in the presence of a weak seed injected into the down-conversion mode. The pump remains a classical field, while the input state is a coherent state $\ket{\beta}$, so that
%%%%%
\begin{equation}\label{Seeded state}
	\ket{\Psi_{wsPDC}} = \hat{U}(t)\,\ket{\beta},
\end{equation}
%%%%%
with
%%%%%
\begin{equation}
	\ket{\beta}= e^{-|\beta|^2/2}\,\sum_{n=0}^{\infty} \frac{\beta^n}{\sqrt{n!}} \ket{n}.
\end{equation}
%%%%%

In the weak-seed regime, $|\beta|^2 \ll 1$, the state can be truncated to
%%%%%
\begin{equation}\label{eq:seed_state}
	\ket{\beta}\simeq \ket{0}+\beta \ket{1},
\end{equation}
%%%%%
where higher Fock components are negligible.

We now expand the evolution operator in the low-gain regime, $|\eta t|\ll 1$, as
%%%%%
\begin{equation}
	\hat U(t) \simeq 1 - \frac{i}{\hbar}\hat H_I t,
\end{equation}
%%%%%
and apply it to the input state. Using $\hat H_I = \hbar \eta\, \hat a^\dagger \hat a^\dagger + \text{H.c.}$ and retaining only terms to first order in $\gamma \equiv \eta t$ and $\beta$, we obtain
%%%%%
\begin{align}
	\ket{\Psi_{wsPDC}} &\simeq \left(1 - i \eta t\, \hat a^\dagger \hat a^\dagger \right)\left(\ket{0} + \beta \ket{1}\right) \nonumber \\
&\simeq \ket{0} + \beta \ket{1} + \gamma \sqrt{2}\,\ket{2} + \gamma \beta \sqrt{6}\,\ket{3}.
\end{align}
%%%%%

Discarding the vacuum and single-photon components, which do not contribute to coincidence measurements, the relevant part of the state reduces to
%%%%%
\begin{equation}\label{eq:seeded_state}
	\ket{\Psi_{wsPDC}} \simeq \gamma \left(\sqrt{2}\,\ket{2} + \sqrt{6}\,\beta \ket{3}\right).
\end{equation}
%%%%%
The first term corresponds to spontaneous pair generation, while the second term arises from the nonlinear interaction in the presence of the seed, yielding a three-photon contribution. This expression captures the single-mode limit of weakly seeded PDC and is closely related to photon-added coherent-state generation~\cite{AgarwalTara1991}.

\subsection{Two- and three-photon coincidence rates}

The third-order normally ordered Glauber correlation function is defined as
%%%%%%
\begin{equation}\label{eq:G3_def}
    G^{(3)}(0,0)
    = \bra{\Psi} \hat{a}^\dagger\hat{a}^\dagger\hat{a}^\dagger
      \hat{a}\hat{a}\hat{a}\ket{\Psi}.
\end{equation}
%%%%%%

Evaluating Eqs.~\eqref{eq:Glauber} and \eqref{eq:G3_def} for the state in Eq.~\eqref{eq:seeded_state}, and retaining only the leading terms in $\gamma$ and $\beta$, we obtain in the weak-seed, low-gain regime
%%%%%%
\begin{equation} \label{eq:G2_seed}
    G^{(2)}_{\mathrm{seed}}(0) \simeq 4|\gamma|^2,
\end{equation}
%%%%%%
and
\begin{equation} \label{eq:G3_seed}
G^{(3)}_{\mathrm{seed}}(0,0) \simeq 36 |\gamma|^2 |\beta|^2.
\end{equation}
%%%%%%

The three-photon contribution therefore scales as $\sim \gamma^2 \beta^2$, reflecting the fact that it originates from pair generation stimulated by a single seed photon. To place this contribution in context, it is useful to compare it with other processes that can lead to three-fold detection events. Spontaneous double-pair emission scales as $\sim \gamma^4$ and is independent of the seed, while the coherent seed alone produces Poissonian three-fold events scaling as $\sim \beta^6$. For $\gamma \ll 1$, this leads to a parameter window
%%%%%%
\begin{equation}\label{eq:narrow window}
\gamma^2 \ll \beta^2 \ll \gamma,
\end{equation}
%%%%%%
in which the seed-induced contribution dominates over multi-pair SPDC while remaining below the intrinsic three-fold background of the coherent field.

An additional distinguishing feature is temporal localization: the seed-induced contribution is concentrated around $(\tau_1,\tau_2)=(0,0)$, whereas three-fold events arising from the seed alone are uniformly distributed in delay space. In practice, this allows operation even for $\beta^2 > \gamma$, provided the signal remains above the noise floor, leading to an effective operating range
%%%%%%
\begin{equation} \label{eq:extended window}
    \gamma^2 \ll \beta^2 \ll 1,
\end{equation}
%%%%%%
while still suppressing multi-pair contributions.

Although the analysis above is based on a single-mode model, the experimentally measured three-fold histogram also includes accidental coincidences. These arise when a genuine two-fold event in one detector pair overlaps with an uncorrelated detection in a third channel. For a coincidence bin of width $\Delta T$, the accidental contribution can be estimated as the product of a two-fold probability and an independent single-click probability. If the singles are dominated by the weak seed, the latter scales as $p_{\mathrm{s}} \simeq S_{\mathrm{s}} \Delta T \propto |\beta|^2$. Using Eq.~\eqref{eq:G2_seed}, this gives
%%%%%%
\begin{equation}\label{eq:G3_a}
    G^{(3)}_{a}(0,0)
    \approx p_{\mathrm{s}}\,G^{(2)}_{\mathrm{seed}}(0)
    \simeq (S_{\mathrm{s}}\Delta T)\,4|\gamma|^{2}
    \ \propto\ 4|\gamma|^{2}|\beta|^{2}
    \ \sim\ \gamma^{2}\beta^{2},
\end{equation}
%%%%%%
showing that accidental contributions scale in the same way as the signal and therefore cannot be neglected. This motivates the normalization procedure introduced in the next section and detailed in the Supplementary Information.

%%%%%%%%%
\begin{figure*}
    \centering
    \includegraphics[width=0.95\linewidth]{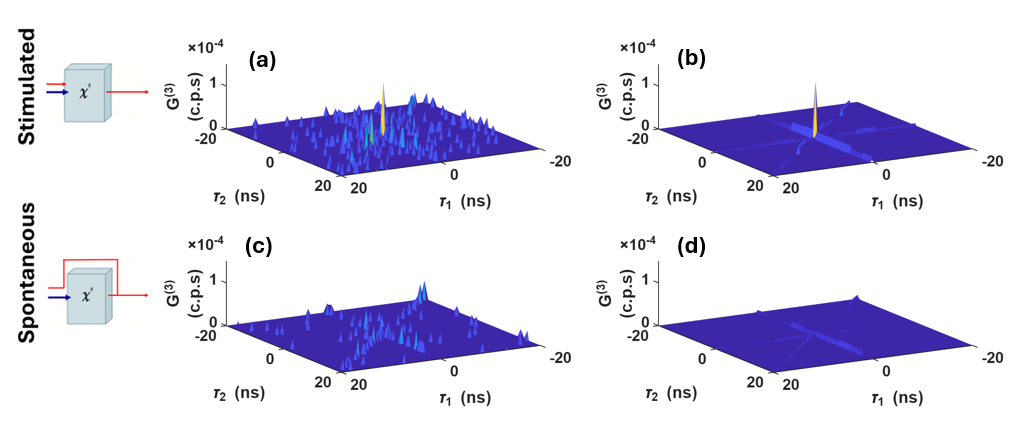} 
    \caption{Pre-normalized three-fold correlations. Top row: (a) measured three-fold coincidence histogram $G^{(3)}(\tau_{1},\tau_{2})$ in the central temporal window with the weak seed present; (b) the same data after block-averaging, which suppresses noise while preserving the underlying coincidence structure. Bottom row: (c) simulated three-fold histogram obtained using the same singles and two-fold rates, but assuming that the seed does not participate in the nonlinear interaction; (d) corresponding block-averaged simulation, showing only the horizontal, vertical, and diagonal features arising from accidental overlaps, with a significantly weaker central feature. }
    \label{fig:exp}
\end{figure*}
%%%%%%%%%
%==========================================================
\section{Experimental setup}
%==========================================================
The experimental setup is shown in Fig.~\ref{fig:setup}. As a light source, we use a tunable, high-power external-cavity diode laser (TA Pro, TOPTICA Photonics) operating in continuous-wave mode at a central wavelength of $800\,\mathrm{nm}$, with a linewidth of $\sim 1\,\mathrm{kHz}$ (FWHM) and an output power of several hundred milliwatts. The beam is expanded and focused into a $15$-mm-long monolithic periodically poled KTP (PPKTP) cavity~\cite{Deng2013BlueKTP}, with its wavefront matched to the cavity mode. The cavity is temperature-stabilized at $65.92\,^\circ\mathrm{C}$ to satisfy the phase-matching condition for second-harmonic generation (SHG), yielding approximately $10\,\mathrm{mW}$ of light at $400\,\mathrm{nm}$, corresponding to an SHG efficiency of about $10\%$. The remaining $800\,\mathrm{nm}$ light, after the attenuation, is used as the seed field.\newline
The generated $400\,\mathrm{nm}$ beam pumps a $1$-mm-thick BBO crystal cut for type-I SPDC in a collinear, near-degenerate configuration at $800\,\mathrm{nm}$. The down-converted photons are spectrally filtered using dielectric band-pass filters with a $10\,\mathrm{nm}$ bandwidth, defining the effective spatio-temporal mode. Together with single-mode optical fiber coupling, this filtering selects a well-defined mode of the field, providing an effective single-mode description for the detected photons. A combination of a half-wave plate (HWP) and a polarizing beam splitter (PBS) allows continuous control over the splitting ratio of the photons between output channels. The light is then coupled into single-mode fibers. One of the output arms is further split using a $50{:}50$ fiber beam splitter, enabling access to all pairwise channel combinations (1-2, 1-3, and 2-3). By adjusting the HWP, the relative detection rates in each channel can be tuned. The total detected SPDC pair rate across all channels is on the order of $10$ pairs/s, ensuring operation in the low-gain regime.\newline
The residual $800\,\mathrm{nm}$ beam is attenuated and spectrally filtered using dichroic mirrors and short-pass filters to form a weak coherent seed. This seed is mode-matched to one of the SPDC collection modes, ensuring spatial and spectral overlap with the down-converted photons. The three output channels are directed to single-photon avalanche detectors (APDs). Detection events are recorded using a Swabian Instruments time-tagger with a temporal bin width of $500\,\mathrm{ps}$. From these data, we reconstruct the full third-order correlation function $G^{(3)}(\tau_{12},\tau_{13})$ by evaluating all pairwise and triple time delays between the detection channels. 

With the pump blocked, the seed produces a detected count rate of approximately $ S_{\mathrm{s}}=10^4$ photons/s. Using the $500\,\mathrm{ps}$ analysis bins, this corresponds to a seed detection probability of $S_s \Delta T \propto 10^{-6}$ per bin, ensuring operation deep in the weak-field regime. Equivalently, the mean seed occupation per filtered optical coherence time is also far below unity: the $10\,\mathrm{nm}$ spectral bandwidth around $800\,\mathrm{nm}$ corresponds to $\Delta\nu\sim 5\times10^{12}\,\mathrm{Hz}$, or $\tau_c\sim 0.2\,\mathrm{ps}$, giving $S_s\tau_c\sim10^{-9}\ll1$.

As discussed in the theoretical section, the measured three-fold histogram includes both genuine correlations and accidental coincidences arising from uncorrelated detection events. To account for this, we normalize the measured three-fold coincidence function $G^{(3)}(\tau_{12},\tau_{13})$ by an estimated accidental background $G^{(3)}_{a}(\tau_{12},\tau_{13})$, constructed from the measured singles and two-fold correlations (see Supplementary Information). This yields the dimensionless quantity
%%%%%%%%
\begin{equation}    \label{eq:g3n_def}
    g^{(3)}_{n}(\tau_{12},\tau_{13})
    = \frac{G^{(3)}(\tau_{12},\tau_{13})}{G^{(3)}_{a}(\tau_{12},\tau_{13})}
    = 1 + \frac{G^{(3)}_{\mathrm{seed}}}{G^{(3)}_{a}},
\end{equation}
%%%%%%%%
which isolates the excess three-photon contribution above the accidental background.

By construction, $g^{(3)}_{n} \rightarrow 1$ in the absence of genuine three-photon correlations. In contrast, when seed-induced contributions are present, Eqs.~\eqref{eq:G3_seed} and \eqref{eq:G3_a} predict an enhancement at zero delay,
%%%%%%%%
\begin{equation}
    g^{(3)}_{n}(0,0)
    \approx 1 + \frac{36|\gamma|^2|\beta|^2}{4|\gamma|^2|\beta|^2}
    = 10,
    \label{eq:g3n_peak}
\end{equation}
%%%%%%%%
providing a clear experimental signature of three-photon correlations. We emphasize that $g^{(3)}_{n}$ is not the standard intensity-normalized Glauber correlation function, but rather a background-normalized quantity constructed to highlight deviations from accidental statistics.\newline

To validate this procedure, we perform a Monte Carlo simulation in which the seed field does not interact with the nonlinear process, so that only accidental coincidences are present. The simulation incorporates the measured singles and two-fold rates, as well as detector timing jitter and dead time. The results, presented in the Supplementary Information, provide a reference for the expected behavior of the system in the absence of seed-induced correlations.

%==========================================================
\section{Results and discussion}
%==========================================================
The measured three-fold coincidence data are summarized in Fig.~\ref{fig:exp}. The top row shows the experimental histograms acquired in the presence of the weak seed, while the bottom row presents a Monte Carlo reference in which the seed is detected but does not participate in the nonlinear interaction. In this simulated case, only spontaneous SPDC and accidental coincidences contribute, allowing us to isolate the role of seed-induced effects.
%%%%%%%%%%
\begin{figure*}
    \centering
    \includegraphics[width=0.95\linewidth]{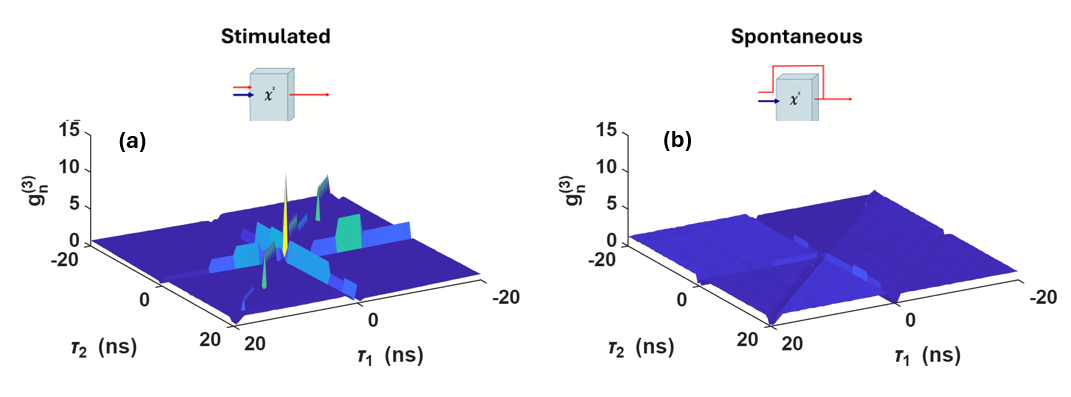}
    \caption{Normalized three-fold correlations. (a) Normalized three-fold function $g^{(3)}_n(\tau_{1},\tau_{2})$ for the experimental data with the weak seed present. A clear peak appears at $\tau_{1}\approx 0$ and $\tau_{2}\approx 0$, while the surrounding region remains close to unity. (b) Corresponding normalized function obtained from the Monte Carlo simulation, including only accidental three-folds and pairwise SPDC correlations. The map remains flat, with $g^{(3)}_n \approx 1$, indicating that the peak observed in (a) cannot be explained by accidentals alone.}
    \label{fig:exp norm}
\end{figure*}
%%%%%%%%%%

Figure~\ref{fig:exp}(a) shows the raw three-fold coincidence function $G^{(3)}(\tau_{1},\tau_{2})$ over the central temporal window. A feature is visible around $\tau_{1}\approx 0$ and $\tau_{2}\approx 0$, but the overall event rate is low, and the signal is largely masked by shot noise. To improve visibility while preserving the underlying temporal structure, we apply a block-averaging procedure that combines neighboring bins without distorting the coincidence features (see Supplementary Information). The processed data, shown in Fig.~\ref{fig:exp}(b), reveal the expected horizontal, vertical, and diagonal ridges associated with accidental overlaps between detector pairs. These structures arise from genuine two-fold coincidences combined with uncorrelated detections in the third channel. Superimposed on these ridges, a localized enhancement at the origin becomes clearly discernible. \newline
The corresponding simulation results are shown in Fig.~\ref{fig:exp}(c,d). Using the same singles and two-fold rates as the experiment, the simulation reproduces the accidental structures with high fidelity. In particular, the horizontal, vertical, and diagonal ridges intersect at the origin, producing a small central feature. However, this feature remains significantly weaker than the peak observed in the experimental data. This comparison indicates that the central enhancement cannot be explained by accidental coincidences alone, nor by higher-order spontaneous processes captured in the simulation.\newline
To quantify this excess contribution, we evaluate the normalized three-fold function $g^{(3)}_{n}$ defined in Eq.~\eqref{eq:g3n_def}, shown in Fig.~\ref{fig:exp norm}. This normalization removes the expected accidental background by construction, allowing any genuine three-photon contribution to appear as a deviation above unity. The experimental data (Fig.~\ref{fig:exp norm}(a)) exhibit a pronounced peak at zero delay, whereas the simulation (Fig.~\ref{fig:exp norm}(b)) remains flat and close to one across the entire delay window. Under identical statistical conditions, a fluctuation of this magnitude would arise from accidental processes with probability $p\approx 10^{-4}$ assuming Poissonian statistics. The observed enhancement, therefore, reflects a genuine three-photon contribution that appears only in the presence of the seed.
A natural interpretation of this feature is that the seed photon participates directly in the nonlinear interaction and stimulates the emission of a signal--idler pair within the same temporal window. In this picture, the seed photon and the generated pair form a temporally correlated triplet, leading to a localized enhancement in the third-order correlation function. This behavior is qualitatively distinct from both spontaneous multi-pair emission and classical-seed amplification, as it arises in a regime where the seed contains, on average, much less than one photon per coherence time.
The measured value $g^{(3)}_{n}(0,0)=12.5$ is in good agreement with the theoretical estimate in Eq.~\eqref{eq:g3n_peak}. Using the measured SPDC rate $R_{S}\sim 10$ pairs/s and the bin width $\Delta T=0.5$~ns, we estimate $\gamma^{2}\approx R_{S}\Delta T/4 \approx 10^{-9}$. From the ratio of Eqs.~\eqref{eq:G2_seed} and \eqref{eq:G3_seed}, we obtain $\beta^{2}\approx G^{(3)}_{\mathrm{seed}}(0,0)/(9\,G^{(2)}_{\mathrm{seed}}(0))\approx 10^{-6}$. These values place the experiment within the parameter regime where the stimulated contribution scales as $\gamma^{2}\beta^{2}$ and dominates over both spontaneous double-pair emission ($\sim \gamma^{4}$) and seed-only background ($\sim \beta^{6}$), consistent with the scaling arguments presented in the theoretical section.

Taken together, these results show that even an ultra-weak coherent seed can induce a measurable modification of the third-order temporal correlations in PDC. The effect manifests as a localized enhancement in $G^{(3)}$, directly linking the presence of the seed to a three-photon correlation that is absent in the purely spontaneous case. This interpretation is supported by the experimental operating regime. The seed field corresponds to an average photon number $\beta^{2}\approx 10^{-6}\ll1$ , such that the interaction is dominated by single-photon components of the coherent state. In this limit, classical-field descriptions based on macroscopic stimulation are not applicable, and the observed scaling and temporal localization of the three-fold signal are consistent with stimulation occurring at the level of individual quanta rather than a classical intensity-driven process.

More broadly, these observations point to a route for engineering higher-order quantum correlations using weakly seeded nonlinear interactions. Unlike approaches based on increased gain or cascaded processes, the present scheme operates in a regime where the nonlinear interaction remains intrinsically low-gain while still exhibiting clear multi-photon signatures. This allows isolating specific contributions without rapid scaling of unwanted higher-order terms.

In addition to their fundamental interest, these results suggest potential applications in correlation-based imaging, spectroscopy, and sensing, where higher-order correlations can provide enhanced sensitivity to temporal and phase-dependent effects~\cite{hodgman2019higher}. The presence of a coherent seed provides a well-defined phase reference, enabling the encoding of phase information in third-order correlations. This could enable interferometric schemes based on photon triplets and provide new tools for probing weak optical effects~\cite{GhobadiInPrepQuantumSeededPDC}. With suitable extensions, such as polarization-entangled pair sources combined with controlled seeding, the same mechanism could be used to generate more complex multipartite states, including three-photon GHZ states, within a single nonlinear platform~\cite{pan2000experimental}.
%===================================================
\section*{Acknowledgments}
%===================================================
Y.K.\ thanks Ms. Achinoam Klein for assistance with the sketches of Fig. 1.
%==========================================================
\section*{Funding}
%==========================================================
This work was supported by the Canada Research Chairs (CRC), Quantum Enhanced Sensing and Imaging (QuEnSI) Alliance Consortia
Quantum grant, and the NRC Quantum Sensing Programme. E.C. was supported by the European Union’s Horizon Europe research and innovation programme under grant agreement No. 101178170, by the Israel Science Foundation under grant agreement No. 2208/24. Y.K. thanks the Israel Ministry of Science for supporting his work on this project through the postdoctoral scholarship program. 

%==========================================================
% Bibliography
%==========================================================

\bibliographystyle{apsrev4-2}
\bibliography{3_fold_bib}

\end{document}

\maketitle
	
	\section{Construction of the three-fold normalization from measured two-fold correlations}
	\label{sec:SI_normalization}
	
	In this section we explain how we construct a model for the expected
	three-fold coincidence rate $G_a^{(\mathrm{3})}(\tau_{12},\tau_{13})$
	using only experimentally measured singles and pairwise coincidence
	histograms. This model describes all three-fold events that can be
	explained by \emph{classical} statistics of pair correlations and
	uncorrelated background. Dividing the measured three-fold histogram
	by this model then isolates any excess three-photon contribution.
	
	\subsubsection*{Basic detection model and notation}
	
	We consider three detectors, labeled $i=1,2,3$, each monitoring a
	stationary optical field. The single-count rate at detector $i$ is
	denoted by
	\begin{equation}
		R_i \quad [\mathrm{s}^{-1}].
	\end{equation}
	In the data acquisition, detection times are binned with a time
	resolution (bin width)
	\begin{equation}
		\Delta T,
	\end{equation}
	so that coincidences are recorded in discrete time bins of width
	$\Delta T$.
	
	For a pair of detectors $(i,j)$ we measure a time-delay histogram
	$H_{ij}(k)$, where the integer index $k$ corresponds to a time delay
	\begin{equation}
		\tau = k \,\Delta T.
	\end{equation}
	The total measurement time is $T$, so that the coincidence \emph{rate
		per bin} is
	\begin{equation}
		G^{(2)}_{ij}(\tau_k)
		=
		\frac{H_{ij}(k)}{T}
		\qquad
		[\mathrm{s}^{-1}\ \text{per bin}].
	\end{equation}
	Analogously, for three detectors we measure a two-dimensional
	histogram $H_{123}(k,\ell)$ as a function of the two delays
	$\tau_{12}=k\Delta T$ and $\tau_{13}=\ell\Delta T$, and define the
	three-fold \emph{rate per bin}
	\begin{equation}
		G^{(3)}(\tau_{12,k},\tau_{13,\ell})
		=
		\frac{H_{123}(k,\ell)}{T}.
	\end{equation}
	
	\subsubsection*{\subsubsection*{Background and correlated contributions in \(G^{(2)}\)}}
	
	Even in the absence of physical correlations between photons, finite single-channel rates $R_i$ and $R_j$ produce a flat accidental coincidence background in the pair histogram. In a simple classical model of independent detection events, the accidental contribution to $G_{ij}^{(2)}(\tau)$ is:
	
	\begin{equation}
		{G^{(2)}_{ij}}_{\mathrm{bg}} = R_i R_j \,\Delta T,
	\end{equation}
	
	which remains independent of $\tau$. For the subsequent calculations, it is necessary to isolate this background contribution from the correlated pairs. We therefore decompose the measured pair histogram into a flat background component and a correlated component:
	
	\begin{equation}
		G^{(2)}_{ij}(\tau) = {G^{(2)}_{ij}}_{\mathrm{bg}} + {G^{(2)}_{ij}}_{\mathrm{corr}}(\tau),
	\end{equation}
	
	where:
	
	\begin{equation}
		{G^{(2)}_{ij}}_{\mathrm{corr}}(\tau) = G^{(2)}_{ij}(\tau) - R_i R_j \,\Delta T.
	\end{equation}
	
	Ignoring statistical noise, ${G^{(2)}_{ij}}_{\mathrm{corr}}(\tau)$ is non-zero only in the narrow region around $\tau=0$ where true physical correlations are present. For large $|\tau|$, we find that $G^{(2)}_{ij}(\tau) \approx R_i R_j \Delta T$, and thus ${G^{(2)}_{ij}}_{\mathrm{corr}}(\tau) \approx 0$. This subtraction ensures that the correlated component vanishes in regions where no physical coincidence peak exists, preventing it from being counted multiple times in the following derivations.
	
	\subsubsection*{Model for the three-fold rate $G_a^{(3)}$}
	
	We now construct a model for the three-fold rate
	$G_a^{(3)}(\tau_{12},\tau_{13})$ that includes all contributions
	arising from \emph{pairwise} correlations and uncorrelated background,
	but excludes any genuine three-photon effect. Three processes contribute:
	
	\begin{enumerate}
		\item A true pair between detectors $(1,2)$ at delay $\tau_{12}$,
		plus an uncorrelated detection at detector~3.
		\item A true pair between detectors $(1,3)$ at delay $\tau_{13}$,
		plus an uncorrelated detection at detector~2.
		\item A true pair between detectors $(2,3)$ at delay
		$\tau_{23} = \tau_{13} - \tau_{12}$, plus an uncorrelated
		detection at detector~1.
	\end{enumerate}
	
	In addition, there is a purely accidental contribution in which all
	three detections are uncorrelated.
	
	\paragraph{(i) Pair $(1,2)$ plus random $3$.}
	The correlated pair rate between detectors~1 and~2 at delay
	$\tau_{12}$ is ${G^{(2)}_{12}}_{\mathrm{corr}}(\tau_{12})$.
	If a third, independent detection at detector~3 occurs with rate
	$R_3$, the corresponding three-fold contribution is
	\begin{equation}
		G^{(3)}_{12+3}(\tau_{12},\tau_{13})
		=
		R_3\, {G^{(2)}_{12}}_{\mathrm{corr}}(\tau_{12})\, \Delta T.
	\end{equation}
	This term depends only on $\tau_{12}$ and is constant along
	$\tau_{13}$.
	
	\paragraph{(ii) Pair $(1,3)$ plus random $2$.}
	Similarly, a correlated pair between detectors~1 and~3 at delay
	$\tau_{13}$, together with an uncorrelated detection at detector~2
	with rate $R_2$, gives
	\begin{equation}
		G^{(3)}_{13+2}(\tau_{12},\tau_{13})
		=
		R_2\, {G^{(2)}_{13}}_{\mathrm{corr}}(\tau_{13})\, \Delta T.
	\end{equation}
	This contribution depends only on $\tau_{13}$.
	
	\paragraph{(iii) Pair $(2,3)$ plus random $1$.}
	For detectors~2 and~3, the relevant delay is
	\begin{equation}
		\tau_{23} = \tau_{13} - \tau_{12},
	\end{equation}
	so that the correlated pair rate is
	${G^{(2)}_{23}}_{(\mathrm{corr})}(\tau_{23})$. Together with an independent
	detection at detector~1 (rate $R_1$), this yields
	\begin{equation}
		G^{(3)}_{23+1}(\tau_{12},\tau_{13})
		=
		R_1\, {G^{(2)}_{23}}_{\mathrm{corr}}(\tau_{23})\, \Delta T
		=
		R_1\, {G^{(2)}_{23}}_{\mathrm{corr}}(\tau_{13}-\tau_{12})\, \Delta T.
	\end{equation}
	Geometrically, this contribution is concentrated along the diagonal
	$\tau_{23} = 0$, i.e.\ $\tau_{13} \approx \tau_{12}$.
	
	\paragraph{(iv) Fully uncorrelated triple background.}
	Even when all three detection events are statistically independent,
	there is a nonzero probability that they fall within the same
	three-fold coincidence bin. In the discrete-time model, the fully
	accidental three-fold rate is
	\begin{equation}
		G^{(3)}_{\mathrm{bg}}
		=
		R_1 R_2 R_3 \, (\Delta T)^2.
	\end{equation}
	
	Collecting all contributions, we obtain the model three-fold rate
	\begin{equation}
		G_a^{(3)}(\tau_{12},\tau_{13})
		=
		G^{(3)}_{12+3}(\tau_{12},\tau_{13})
		+  G^{(3)}_{13+2}(\tau_{12},\tau_{13})
		+  G^{(3)}_{23+1}(\tau_{12},\tau_{13})
		+ G^{(3)}_{\mathrm{bg}},
	\end{equation}
	i.e.
	\begin{equation}
		\begin{aligned}
			&G_a^{(3)}(\tau_{12},\tau_{13})
			=
			R_3\, {G^{(2)}_{12}}_{\mathrm{corr}}(\tau_{12})\, \Delta T
			+ R_2\, {G^{(2)}_{13}}_{\mathrm{corr}}(\tau_{13})\, \Delta T \\
			&\quad
			+ R_1\, {G^{(2)}_{23}}_{\mathrm{corr}}(\tau_{13}-\tau_{12})\, \Delta T
			+ R_1 R_2 R_3 \, (\Delta T)^2.
		\end{aligned}
		\label{eq:SI_G3a_model}
	\end{equation}
	
	In regions of $(\tau_{12},\tau_{13})$ where no physical pair
	correlations are present, ${G^{(2)}_{ij}}_{\mathrm{corr}}(\tau) \approx 0$, and the model reduces
	to the flat background
	\begin{equation}
		G_a^{(3)}(\tau_{12},\tau_{13})
		\approx
		R_1 R_2 R_3 \, (\Delta T)^2.
		\label{eq:background_G3}
	\end{equation}
	When all three-fold events are explained by independent counting
	statistics and pairwise correlations only, the experimental three-fold
	histogram $G_3$ should show the same behavior, and the ratio
	\begin{equation}
		g^{(3)}_{n}(\tau_{12},\tau_{13})
		= \frac{G^{(3)}(\tau_{12},\tau_{13})}{G^{(3)}_{a}(\tau_{12},\tau_{13})}
		\approx 1
		\label{eq:SI_normalized_g3}
	\end{equation}
	Therefore, values of $g^{(3)}_{n}(\tau_{12},\tau_{13})$ significantly larger than unity indicate an \emph{excess}
	of three-fold coincidences beyond what is expected from the measured
	two-fold correlations and singles rates.

%%%%%%%%%%%%%%%%%%%%%%%%%%%%%%%%%%%%%%%%%%%%%%%%
\section{Treatment and coarse-graining of the experimental three-fold histogram}
\label{sec:SI_pixelating}

The time tagger records three-fold time differences for the different detector-channel permutations. The corresponding histograms are first accumulated separately and then rotated and shifted into a common reference frame with axes \(\tau_{12}\) and \(\tau_{13}\). This yields a symmetric experimental map \(G^{(3)}_{\mathrm{raw}}(\tau_{12},\tau_{13})\) that contains the full three-fold coincidence information.

For the analysis, we retain only the central window around zero delay,
\begin{equation}
    |\tau_{12}| \leq \tau_{\mathrm{max}}, \qquad
    |\tau_{13}| \leq \tau_{\mathrm{max}},
\end{equation}
that is, the central square block of the full histogram. This region contains the central three-fold peak and its immediate surroundings.

\subsubsection*{Coarse-graining}

Because the raw two-dimensional histogram contains limited counts per bin, especially away from the central peak, we apply two complementary coarse-graining steps. First, we average over large background-dominated regions of the 2D map. Second, we average the data along the three coincidence lines over finite time pixels, where the signal is concentrated. This suppresses statistical noise while preserving the narrow coincidence features.

In the \(\tau_{12}\)-\(\tau_{13}\) plane, the coincidence signal is concentrated along the three lines
\[
\tau_{12}=0, \qquad \tau_{13}=0, \qquad \tau_{23}=0,
\]
with
\[
\tau_{23}=\tau_{13}-\tau_{12}.
\]
Away from these lines and away from the center, the histogram is dominated by accidental triples and varies only slowly. We therefore divide the central 2D block into several background regions \(\mathcal{R}_\alpha\): two off-diagonal quadrants and four triangular regions near the diagonals, excluding the central strip. For each region we compute the mean count
\begin{equation}
    \langle G^{(3)}_{\mathrm{raw}} \rangle_\alpha
    =
    \frac{1}{|\mathcal{R}_\alpha|}
    \sum_{(\tau_{12},\tau_{13}) \in \mathcal{R}_\alpha}
    G^{(3)}_{\mathrm{raw}}(\tau_{12},\tau_{13}),
\end{equation}
where \(|\mathcal{R}_\alpha|\) is the number of pixels in \(\mathcal{R}_\alpha\). Away from the coincidence lines, the coarse-grained map is then defined by
\begin{equation}
    G^{(3)}(\tau_{12},\tau_{13})
    =
    \langle G^{(3)}_{\mathrm{raw}} \rangle_\alpha,
    \qquad
    (\tau_{12},\tau_{13}) \in \mathcal{R}_\alpha.
\end{equation}

Near the coincidence lines, a higher-resolution treatment is required. We therefore extract three one-dimensional cross sections from the central 2D block,
\begin{align}
    C_{12|3}(\tau) &= G^{(3)}_{\mathrm{raw}}(0,\tau), \\
    C_{13|2}(\tau) &= G^{(3)}_{\mathrm{raw}}(\tau,0), \\
    C_{23|1}(\tau) &= G^{(3)}_{\mathrm{raw}}(\tau,\tau).
\end{align}
These correspond to the two axes through the origin and to the main diagonal. Each cross section contains a narrow coincidence peak on top of a slowly varying accidental background.

To reduce the noise, each cross section is rebinned into non-overlapping time pixels of width
\begin{equation}
    T_{\mathrm{pix}} > \Delta T,
\end{equation}
where \(\Delta T\) is the original bin width. The averaging is performed only for \(\tau \neq 0\), namely separately in the regions \(\tau<0\) and \(\tau>0\), so that the central bin at \(\tau=0\) is excluded from the averaging. For a generic cross section \(C(\tau)\), we define the averaged pixel value
\begin{equation}
    \bar{C}_m
    =
    \frac{1}{N_m}
    \sum_{\substack{\tau \in \text{pixel } m \\ \tau \neq 0}} C(\tau),
\end{equation}
where \(N_m\) is the number of original time bins contained in pixel \(m\), and assign to that pixel the average delay \(\bar{\tau}_m\). This averaging over \(T_{\mathrm{pix}}/\Delta T\) original bins reduces statistical fluctuations. Applying this procedure to the three cross sections yields the pixelated profiles
\(\bar{C}_{12|3}(\bar{\tau}_m)\),
\(\bar{C}_{13|2}(\bar{\tau}_m)\), and
\(\bar{C}_{23|1}(\bar{\tau}_m)\).

The values of the coarse-grained map along the three coincidence lines are then taken from these averaged profiles. For \(\tau \neq 0\) belonging to pixel \(m\), we assign
\begin{align}
    G^{(3)}(0,\tau) &= \bar{C}_{12|3}(\bar{\tau}_m), \\
    G^{(3)}(\tau,0) &= \bar{C}_{13|2}(\bar{\tau}_m), \\
    G^{(3)}(\tau,\tau) &= \bar{C}_{23|1}(\bar{\tau}_m).
\end{align}
Thus, along the coincidence lines the coarse-grained map is obtained by averaging over each time pixel for all \(\tau \neq 0\). At the central point, corresponding to the three-fold coincidence peak, we keep the experimental value unchanged,
\begin{equation}
    G^{(3)}(0,0) = G^{(3)}_{\mathrm{raw}}(0,0).
\end{equation}
Together with the regional averages used away from the coincidence lines, this defines the experimentally reconstructed and coarse-grained three-fold rate \(G^{(3)}\), which is used in Eq.~\eqref{eq:SI_normalized_g3}.
\clearpage 

%%%%%%%%%%%%%%%%%%%%%%%%%%%%%%%%%%%%%%%%%%%%%%%%
\section{Monte-Carlo simulation of the three-fold measurement}
\label{sec:SI_simulation}

To validate the analysis protocol and estimate the sensitivity of the experiment, we perform a Monte-Carlo simulation of the three-fold measurement. In this simulation, we explicitly assume that the seed field does \emph{not} modify the statistics of the SPDC source, i.e., no stimulated three-photon process is present. All simulated three-fold events therefore arise solely from accidental coincidences and pairwise SPDC correlations. This provides a reference case in which the normalized three-fold function \(g_n^{(3)}\) is expected to be unity up to statistical fluctuations.

\subsection*{Model assumptions}

We model the source as a continuous-wave (CW) pumped SPDC crystal that generates photon pairs at a mean rate \(R_{\mathrm{p}}\) (pairs per second). On the detection time scale, this process is well approximated as a stationary Poisson process. Each SPDC event produces exactly one photon pair (signal and idler) with strongly correlated detection times. The temporal correlation between the two photons is described by a Gaussian distribution with effective width
\begin{equation}
    \sigma_{\mathrm{corr}},
\end{equation}
which accounts for both the intrinsic SPDC correlation time and the detector timing jitter.

The optical network downstream of the crystal (beam splitters, fiber couplers, and detector efficiencies) is modeled by three effective detection probabilities
\begin{equation}
    \eta_1, \ \eta_2, \ \eta_3,
\end{equation}
which give the probability that a given photon is detected in channel 1, 2, or 3, respectively. The remaining probability,
\[
1-\eta_1-\eta_2-\eta_3,
\]
corresponds to loss. In addition, each detector \(i\) has an independent background rate \(B_i\), which includes both dark counts and stray light.

\subsection*{Generation of detection times}

We simulate a total acquisition time \(T_{\mathrm{sim}}\) and generate a list of detection times for each of the three channels as follows.

\paragraph{(i) SPDC pair times.}
We first draw the total number of SPDC pairs,
\begin{equation}
    N_{\mathrm{pairs}} \sim \mathrm{Poisson}(R_{\mathrm{p}}T_{\mathrm{sim}}).
\end{equation}
For each pair \(j=1,\dots,N_{\mathrm{pairs}}\), we choose a creation time uniformly in the interval \([0,T_{\mathrm{sim}})\),
\begin{equation}
    t_j \sim \mathcal{U}[0,T_{\mathrm{sim}}).
\end{equation}

\paragraph{(ii) Correlated detection times within a pair.}
For each SPDC pair we generate two detection times,
\begin{equation}
    t^{(A)}_j = t_j + \delta^{(A)}_j, \qquad
    t^{(B)}_j = t_j + \delta^{(B)}_j,
\end{equation}
where the temporal offsets \(\delta^{(A)}_j\) and \(\delta^{(B)}_j\) are independent Gaussian random variables,
\begin{equation}
    \delta^{(A)}_j, \ \delta^{(B)}_j
    \sim \mathcal{N}(0,\sigma_{\mathrm{corr}}^2).
\end{equation}
This produces a narrow coincidence peak in the simulated pair histograms, consistent with the experiment.

\paragraph{(iii) Routing to the three detectors.}
Each photon in the pair is routed independently to one of the three detectors, or lost, according to the probabilities \(\eta_1,\eta_2,\eta_3\). For photon \(A\), for example, we draw a uniform random number \(r^{(A)}_j \in [0,1)\) and assign
\begin{equation}
    \text{detector} =
    \begin{cases}
        1, & r^{(A)}_j < \eta_1, \\
        2, & \eta_1 \le r^{(A)}_j < \eta_1+\eta_2, \\
        3, & \eta_1+\eta_2 \le r^{(A)}_j < \eta_1+\eta_2+\eta_3, \\
        \text{lost}, & r^{(A)}_j \ge \eta_1+\eta_2+\eta_3.
    \end{cases}
\end{equation}
The same rule is applied independently to photon \(B\). This yields three lists of SPDC detection times, \(\{t^{\mathrm{(SPDC)}}_{i,k}\}\), for channels \(i=1,2,3\).

\paragraph{(iv) Background counts.}
For each detector \(i\), we generate an independent Poisson process of background events with mean rate \(B_i\). The number of background events is
\begin{equation}
    N^{\mathrm{(bg)}}_i \sim \mathrm{Poisson}(B_iT_{\mathrm{sim}}),
\end{equation}
and their detection times are drawn uniformly in the interval \([0,T_{\mathrm{sim}})\),
\begin{equation}
    t^{\mathrm{(bg)}}_{i,k} \sim \mathcal{U}[0,T_{\mathrm{sim}}).
\end{equation}
A more refined model based on a phase-diffusing coherent state can also be used, but it does not affect the conclusions.

\paragraph{(v) Final detection streams.}
The complete set of detection times in channel \(i\) is then
\begin{equation}
    \{ t_{i,k} \}
    =
    \{ t^{\mathrm{(SPDC)}}_{i,k} \}
    \cup
    \{ t^{\mathrm{(bg)}}_{i,k} \},
\end{equation}
which is sorted in ascending order. These time streams are then processed using the same time-tag analysis as the experimental data.

\subsection*{Construction of simulated histograms}

Given the three detection streams \(\{t_{1,k}\}\), \(\{t_{2,k}\}\), and \(\{t_{3,k}\}\), we compute the simulated pair and three-fold coincidences using the same time-window logic and binning parameters as in the experiment.

\paragraph{(i) Pair histograms.}
For each detector pair \((i,j)\), we loop over the events in channel \(i\) and search for events in channel \(j\) that fall within the coincidence window \(|\tau|<\tau_{\max}\). For each detected pair we record the delay
\begin{equation}
    \tau_{ij} = t_i - t_j,
\end{equation}
and accumulate a histogram \(H_{ij}(k)\) with bin width \(\Delta T\). Dividing by the total simulation time \(T_{\mathrm{sim}}\) yields the pair rate per bin \(G^{(2)}_{ij}(\tau_k)\) used in the normalization model.

\paragraph{(ii) Three-fold histogram.}
To construct the simulated three-fold histogram, we identify all triplets of detections \((t_1,t_2,t_3)\) such that the delays \(\tau_{12}=t_1-t_2\) and \(\tau_{13}=t_1-t_3\) satisfy
\begin{equation}
    |\tau_{12}| < \tau_{\max}, \qquad |\tau_{13}| < \tau_{\max}.
\end{equation}
For each such triplet, we increment the 2D histogram \(H_{123}(k,\ell)\) in the bin corresponding to \((\tau_{12,k},\tau_{13,\ell})\). The resulting three-fold rate per bin is
\begin{equation}
    G^{(3)}_{\mathrm{sim}}(\tau_{12,k},\tau_{13,\ell})
    =
    \frac{H_{123}(k,\ell)}{T_{\mathrm{sim}}}.
\end{equation}
Up to statistical noise, this synthetic histogram plays the same role as the experimental \(G^{(3)}\) in the analysis.

\subsection*{Simulation-based validation of the normalization protocol}

One purpose of the simulation is to test the normalization procedure introduced in Eq.~\eqref{eq:SI_normalized_g3}. From the simulated pair histograms \(G_{12}(\tau)\), \(G_{13}(\tau)\), and \(G_{23}(\tau)\), together with the singles rates \(R_i\) extracted from the simulated data, we construct the model three-fold rate \({G^{(3)}_a}_{\mathrm{sim}}(\tau_{12},\tau_{13})\) using Eq.~\eqref{eq:SI_G3a_model}. The normalized three-fold function in the simulation is then
\begin{equation}
    {g^{(3)}_n}_{\mathrm{sim}}(\tau_{12},\tau_{13})
    =
    \frac{G^{(3)}_{\mathrm{sim}}(\tau_{12},\tau_{13})}
    {{G^{(3)}_a}_{\mathrm{sim}}(\tau_{12},\tau_{13})}.
\end{equation}

By construction, the simulation contains only accidental coincidences and pairwise SPDC correlations; no genuine three-photon process is present. Therefore,
\begin{equation}
    {g^{(3)}_n}_{\mathrm{sim}}(\tau_{12},\tau_{13}) \approx 1
\end{equation}
for all delays \((\tau_{12},\tau_{13})\). This is indeed observed (Fig.~3 in the main text): although the raw three-fold histogram \(G^{(3)}_{\mathrm{sim}}\) exhibits apparent structure due to overlapping pair correlations, the normalized function \({g^{(3)}_n}_{\mathrm{sim}}\) is flat and consistent with unity within statistical uncertainty.